\documentclass[submission]{eptcs}
\providecommand{\event}{Linearity 2016}

\usepackage{amsmath}

\usepackage{dashrule}

\usepackage{float}
\usepackage{graphicx}

\usepackage{caption}
\floatstyle{plain}
\newfloat{program}{thp}{lop}
\floatname{program}{Program}

\usepackage{latexsym}
\usepackage{amssymb}            % for \multimap (-o)
\usepackage{stmaryrd}           % for \binampersand (&), \bindnasrepma (\paar)
\newcommand{\m}[1]{\mathsf{#1}}
\newcommand{\mi}[1]{\mathit{#1}}

\newcommand{\semi}{\mathrel{;}}

\newcommand{\vvdash}{\mathrel{\vdash\kern-0.8ex\vdash}}
\newcommand{\lolli}{\multimap}
\newcommand{\tensor}{\otimes}

\usepackage[usenames,dvipsnames,svgnames,table]{xcolor}

\usepackage{multirow}

\newcommand{\denote}[1]{\text{$[\![ $#1$ ]\!]$}}

\newcommand{\lft}[1]{\overleftarrow{\mathstrut #1}}
\newcommand{\rgt}[1]{\overrightarrow{\mathstrut #1}}

\title{Non-Blocking Concurrent Imperative Programming with Session Types}
\author{Miguel Silva
\institute{Department of Computer Science\\
University of Porto\\
Porto, Portugal}
\and
M\'{a}rio Florido
\institute{Department of Computer Science\\
University of Porto\\
Porto, Portugal}
\and
Frank Pfenning
\institute{Carnegie Mellon University\\
Pennsylvania, USA}
}

\def\titlerunning{Non-Blocking Concurrent Imperative Programming with Session Types}
\def\authorrunning{M. Silva, M. Florido \& F. Pfenning}

\begin{document}

\maketitle

\begin{abstract}

Concurrent C0 is an imperative programming language in the C family with session-typed message-passing concurrency. The previously proposed semantics implements asynchronous (non-blocking) output; we extend it here with non-blocking input. A key idea is to postpone message reception as much as possible by interpreting receive commands as a request for a message. We implemented our ideas as a translation from a blocking intermediate language to a non-blocking language. Finally, we evaluated our techniques with several benchmark programs and show the results obtained. While the abstract measure of span always decreases (or remains unchanged), only a few of the examples reap a practical benefit.
\end{abstract}

\section{Introduction}

A session describes the collective conduct of the components of a concurrent system. Binary sessions focus on the interactions between two of these components, with an inherent concept of duality. A \textit{session type} \cite{honda1993types, honda1998language} defines the communication between processes using this notion. Session types enforce conformance to a communication protocol, organizing a session to occur over a communication channel. Recently, session types have been linked with linear logic via a Curry-Howard interpretation of linear propositions as types, proofs as processes, and cut reduction as communication. Variations apply for both intuitionistic \cite{caires2010session, caires2013linear} and classical \cite{wadler2012propositions} linear logic. The intuitionistic variant culminated in SILL, a functional language with session-typed concurrency \cite{Toninho2013}.

The adaptation of SILL to other paradigms of programming gave rise to CLOO (Concurrent Linear Object-Orientation) \cite{balzer2015objects}, a concurrent object-oriented language that types objects and channels with session types, and Concurrent C0 \cite{willseydesign}, a session-based extension to an imperative language. In this paper, we present a non-blocking model of receiving messages in Concurrent C0. Our model will be based on the premise that we only want to halt execution to wait for a message when the data contained in that message is necessary to continue computation. We will then present a compilation function from the original version of Concurrent C0 to our non-blocking model. We also present a cost semantics for Concurrent C0 that computes the span and work~\cite{Cormen:2001:IA:580470}, providing an abstract analytical measure of latent parallelism in the computation. The span always improves or remains the same under the non-blocking semantics, although the experimental results show that it is difficult to exploit these improvements in practice.

In related work, Guenot~\cite{Guenot14un} has given a pure and elegant computational interpretation of classical linear logic with non-blocking input based on the solos calculus~\cite{Laneve03mscs}. The primary notion of a process as a thread of control that pervades our work is no longer visible there and it does not immediately suggest an implementation strategy. Our work generalizes the functional notions of \textit{futures} and \textit{promises}~\cite{baker1977futures, Friedman78aspectsof} by supporting more complex bidirectional communication~\cite{Toninho12fossacs}.

Section 2 briefly introduces the Concurrent C0 language, outlining its syntax and operational semantics. Section 3 exposes the model for non-blocking reception of messages. Section 4 defines a translation from the original to the non-blocking language. Section 5 provides an experimental evaluation of the implementation of the new model, and Section 6 concludes.

\section{Imperative Programming with Sessions Types}

This section introduces the main concepts of programming of Concurrent C0 (CC0), a session-based extension of C0 \cite{c0}, itself a small safe subset of C augmented with contracts. For a more in-depth description of CC0, we refer the reader to a technical report \cite{willseydesign}.

A program in CC0 is a collection of processes exchanging messages through channels. These processes are \textit{spawned} by functions that return channels. The process that calls these spawning functions is called the client. The new process at the other end of the channel is called the provider, who is said to \textit{offer a session} over the channel.

CC0's session typing system is based on work by Caires and Pfenning \cite{caires2010session}, who establish a correspondence between session types for $\pi$-calculus and intuitionistic linear logic. CC0's channels have a linear semantics: there is exactly one reference to the channel besides the provider's. This captures the behavior described in the previous paragraph: the process we called the client has the unique reference to the provided channel.

Messages are sent asynchronously: processes advance in parallel without waiting for acknowledgement of the sent message being received. Programmers must specify the protocol of message exchange using session types and a linear type system enforces concordance with this protocol \cite{pfenning2015polarized}.

\begin{program}
\footnotesize
\begin{verbatim}		
choice queue {
  <?int; ?choice queue>  Enq;      // receive 'Enq', then int, continue as 'queue'
  <!choice queue_elem>   Deq;      // receive 'Deq', continue as 'queue_elem'
  <!bool; ?choice queue> IsEmpty;  // receive 'IsEmpty', then send bool, continue as 'queue'
  < >                    Dealloc;  // receive 'Dealloc', then terminate
};
choice queue_elem {
  <?choice queue>       None;      // send 'None', continue as 'queue'
  <!int; ?choice queue> Some;      // send 'Some', send int, continue as 'queue'
};
typedef <?choice queue> queue;    // define 'queue' as external choice
\end{verbatim}
\caption{Protocol definition of a queue with constant time enqueue and dequeue operations from client's perspective, in CC0.}
\label{program:session}
\end{program}
\normalsize

Program \ref{program:session} is an example defining several session types in CC0. We use \verb'<...>' to enclose a session type, \verb'?' to indicate an input and \verb'!' to indicate and output. An empty \verb'< >' represents the end of a session. A \emph{choice} indicates that a label is received (external choice, prefixed by \verb'?') or that a label is sent (internal choice, prefixed by \verb'!'). Choices have to be explicitly named and declared in CC0 in a syntax inspired by structs in C.

% For example, the choice named $\mathit{queue}$ is a choice between labels \verb'Enq', \verb'Deq', \verb'IsEmpty' and \verb'Dealloc'. It is only used after an output prefix (\verb'?'), which means it always represents an external choice, made by the client and sent to the provider of the queue interface.  In contrast, the choice $\mathit{queue\_elem}$ is an internal choice (always prefixed by \verb'!'), which means the provider has to send label \verb'None' (no element in the queue) or \verb'Some' (some element in the queue).

The queue's protocol exchanges messages in two directions, from the client process to the provider process and vice-versa, resulting in each direction being conveyed independently through an external ($queue$) and internal choice ($queue\_elem$). The session type is given from the provider's perspective and the compiler determines statically if the operations indicated in the protocol are performend in the correct order according to the channel's session type. 

CC0 is compiled to a target language and linked with a runtime system, both written in C, responsible for implementing communication.

%The compiler checks if messages are being exchanged in the correct order, in agreement with the session type, and enforces linear use of channels.

The session typing of CC0 uses polarized logic, as detailed in \cite{pfenning2015polarized}, to maintain the direction of the communication. Positive polarity indicates that information is streaming from the provider and negative to the provider. A \textit{shift} is used to swap polarities. The runtime system explicitly tracks the polarity of each channel and the compiler infers and inserts the minimal amount of \textit{shifts} into the target language.

CC0 has the usual features of an imperative programming language, such as conditionals, loops, assignments and functions, extended by communication primitives. Below, we present the core communication syntax of the target language. We differentiate a channel from a variable by placing a $\$$ before the channel name, such as $\$c$. 

\footnotesize
\[
\begin{array}{lcl@{\hspace{1em}}l}
P, Q & ::= & \$c = spawn(P); \; Q & \m{spawn} \\
& \mid & \$c = \$d & \m{forward} \\
& \mid & \m{close}(\$c) & \mbox{send $\m{end}$ and terminate} \\
& \mid & \m{wait}(\$c); \; Q & \mbox{receive $\m{end}$} \\
& \mid & \m{send}(\$c, \; e); \; Q & \mbox{send data (including channels)} \\
& \mid & x = \m{recv}(\$c); \; Q & \mbox{receive data (including channels)} \\
& \mid & \m{send}(\$c, \; \mi{shift}); \; Q & \mbox{send shift} \\
& \mid & \mi{shift} = \m{recv}(\$c); \; Q & \mbox{receive shift} \\
& \mid & \$c.\mi{lab}; \; Q & \mbox{send label} \\
& \mid & \m{switch}(\$c) \; \{\mi{lab}_i \rightarrow P_i\}_i  & \mbox{receive label} \\
\end{array}
\]
\normalsize

\subsection{Cost Semantics}

Pfenning and Griffith introduced asynchronous communication using polarized logic \cite{pfenning2015polarized} in the operational semantics of the target language, which is expressed as a \textit{substructural operational semantics} \cite{Pfenning09lics}, based on \textit{multiset rewriting} \cite{Cervesato09ic}. We now extend this operational semantics by assigning a cost to each operation, using the work-span model \cite{Cormen:2001:IA:580470}. We will only count communication costs, ignoring internal computation. Although this may not lead to an entirely realistic measure for the complexity of an algorithm, it will still make for an interesting abstract one. Moreover, in many of our examples communication costs dominate performance.

In the cost semantics we maintain a span $s$ for each executing process which represents the earliest global time (counting only communication steps) at which the process could have reached its current state. Because messages can only be received after they have been sent, each message is tagged with the time at which it is sent, and the recipient takes the maximum between its own span and the span carried by the message. Except for operations using \textit{shifts} or \textit{forwards}, each call to a communication function will increase the span by one unit.

The work is determined individually by each process. As with span, all operations except the ones using \textit{shifts} or \textit{forwards} will increase work by one unit. Although each message also carries the work of the sending process, this work is ignored unless the message is an $\m{end}$ or \textit{forward}. In these two cases, the receiving process will add the work carried by the message to its own, to propagate the work of the sending process, which is being terminated.

\footnotesize

\[
\begin{array}{llcl}
\mbox{Configurations} & \Omega & ::= & \cdot \\
& & \mid & \m{queue}(\$c, q, \$d), \Omega \\
& & \mid & \m{proc}(\$c, P, s, w), \Omega \\
& & \mid & \m{cell}(\$c, x, v), \Omega \\
\end{array}
\]

\[
\begin{array}{l@{\hspace{1em}}lcll}
\mbox{Queue filled by provider} & \lft{q} & ::= & \lft{\cdot} \mid \lft{(v,s,w)\cdot q} \mid \lft{(\m{end},s,w)} \mid \lft{(\m{shift},s,w)} \\[1ex]
\mbox{Queue filled by client} & \rgt{q} & ::= & \rgt{(\m{shift},s,w)} \mid \rgt{q\cdot (v,s,w)} \mid \rgt{\cdot}
\end{array}
\]

\normalsize

Configurations describe executing processes, message queues connecting processes (one for each channel), and local storage cells.  In our definition, $\m{proc}(\$c, P, s, w)$ is the state of a process executing program P, offering along channel $\$c$, with span $s$ and work $w$. The message queue is represented by $\m{queue}(\$c, q, \$d)$, which connects processes offering along $\$d$ with a client using $\$c$. The memory cell $\m{cell}(\$c, x, v)$ holds the state of variable $x$ with value $v$ in the process offering along channel $\$c$. We use the multiplicative conjunction ($\tensor$) to join processes, queues, and memory cells, and linear implication to express state transition ($\lolli$). Due to space constraints, we show only some representative rules.

\footnotesize

\[
\begin{array}{lcl}
\m{shift}\_\m{s} & : & \m{queue}(\$c, \lft{q}, \$d) \tensor \m{proc}(\$d, \m{send}(\$d, \m{shift}) \semi P, s, w) \\
& & \quad \lolli \{ \m{queue}(\$c, \lft{q \cdot (shift,s,w)}, \$d) \tensor \m{proc}(\$d, P, s, w) \} \\[1ex]
\m{data}\_\m{r} & : & \m{proc}(\$e, x = \m{recv}(\$c) \semi Q, s, w) \tensor
\m{queue}(\$c, \lft{(v,s_1,w_1)\cdot q}, \$d) \\
& & \quad \lolli \{\exists x.\, \m{proc}(\$e, Q, \m{max}(s, s_1) + 1, w+1) \tensor \m{queue}(\$c, \lft{q}, \$d) \tensor \m{cell}(\$e, x, v) \} \\[1ex]
\m{close} & : & \m{queue}(\$c, \lft{q}, \$d) \tensor \m{proc}(\$d, \m{close}(\$d), s,w) \\
& & \quad \lolli \{ \m{queue}(\$c, \lft{q \cdot (\m{end},s+1,w+1)}, \_) \} \\[1ex]
\m{wait} & : & \m{proc}(\$e, \m{wait}(\$c) \semi Q,s,w) \tensor
\m{queue}(\$c, \lft{(\m{end},s_1,w_1)}, \_) \\
& & \quad \lolli \{ \m{proc}(\$e, Q, \m{max}(s,s_1)+1, w + w_1 + 1) \} \\[1ex]
\m{fwd}\_\m{s} & : & \m{queue}(\$c, \lft{q}, \$d) \tensor \m{proc}(\$d, \$d = \$e, s, w) \\
& & \quad \lolli \{ \m{queue}(\$c, \lft{q \cdot (\m{fwd}, s, w)}, \$e) \} \\[1ex]
\m{fwd}\_\m{r} & : & \m{proc}(\$d, P(\$c), s, w) \tensor
\m{queue}(\$c, \lft{(\m{fwd},s_1, w_1)}, \$e) \\
& & \quad \lolli \{ \m{proc}(\$d, P(\$e), \m{max}(s,s_1), w + w_1)  \} \\[1ex]
\m{spawn} & : & \m{proc}(\$c, \$d = P() \semi Q, s, w) \\
& & \quad \lolli \{\exists \$d.\, \m{proc}(\$c, Q, s, w) \tensor \m{queue}(\$c, \lft{\cdot}, \$d) \tensor \m{proc}(\$d, P, s, 0) \}
\end{array}
\]

\normalsize

\section{Non-Blocking Receive}

In this section, we present the main contribution of our work: a new model for message reception whose goal will be to block the execution of the process to wait for a message only when the data contained in this message is necessary to continue the execution. 

Receiving a message in CC0 blocks the execution of the program, a behavior matching the operational semantics. A receiving function ($\m{recv}$ or $\m{wait}$) will only succeed when it is possible to retrieve the corresponding message from the queue of the associated channel.

This model for message reception may not be the optimal choice for some algorithms where an arbitrary imposed order on messages received on two different channels might prevent other computation to proceed. An extreme case is when a received value is not actually ever needed.

Our alternative for follows two principles.  One, the difference should be invisible to the programmer who should not need to know exactly when a message is received. Second, the implementation will still need to adhere to the protocol defined by the session type, which forces the order of sends and receives.

\subsection{Runtime Implementation}
We will now focus on the runtime system, describing how the concepts previously introduced are implemented and how we solved the issues that arose from our model. Our baseline CC0 runtime is implemented in C, using the \textit{pthread} library in a straightforward way to obtain parallelism in multi-core machines. Message passing is implemented using the queues in shared memory.

When a process in CC0 is spawned, it provides a channel to its parent process, i.e., the process that called the spawn function. This channel is represented internally as a structure that aggregates crucial information, namely, a queue of messages.

Our model involves postponing reception as much as possible, by interpreting receives as a request for a message. The request is saved on the channel until a synchronization is necessary. A synchronization is triggered when some data contained on any of the requests is required to continue execution. For example,  if a process requested a $\m{shift}$ from a channel, a synchronization is required to correct the polarity of the channel, that is, drain the message queue in order to change the direction of communication using the same queue.

The requests are handled in the order they were made, which guarantees that the session type is still being respected. Furthermore, all these changes only occur in an intermediate language, allowing CC0 to keep the same source-level syntax.

From a low level implementation point of view, this change required adding a queue data structure to the channel to hold the requests, creating synchronization functions and modifying the existing receiving functions. It was also necessary to define a translation function between the original, blocking, intermediate language to the new, non-blocking, one. This translation is introduced in the next section. 
Here are the new constructs introduced by the translation.

\footnotesize
\[
\begin{array}{lcl@{\hspace{1em}}l}
P, Q & ::= & \ldots \\
& \mid & \m{async\_wait}(\$c); \; Q & \mbox{request an $\m{end}$} \\
& \mid & x = \m{async\_recv}(\$c); \; Q & \mbox{request data} \\
& \mid & \mi{shift} = \m{async\_recv}(\$c); \; Q & \mbox{request a $\m{shift}$} \\
& \mid & \m{sync}(\$c, x); \; Q & \mbox{synchronize variable} \\
& \mid & \m{sync}(\$c, \m{shift}); \; Q & \mbox{synchronize $\m{shift}$} \\
& \mid & \m{sync}(\$c, \m{end}); \; Q & \mbox{synchronize $\m{end}$} \\
\end{array}
\]
\normalsize

Program \ref{program:targetlang} shows a possible implementation for the queue example from the previous section. It is presented using an abbreviated version of the target language, using the non-blocking model. Note that receiving a label requires branching between different cases and therefore continues to block in order to avoid speculative execution.

\begin{program}
\footnotesize
\begin{tabular}{l|l}
\begin{minipage}{7.7cm}
\begin{verbatim}
queue $q elem (int x, queue $r) {
  switch ($q) {
    case Enq:
      int y = async_recv($q);
      $r.Enq; 
      sync($q, y); send($r, y);
      $q = elem(x, $r);
    case Deq:
      shift = async_recv($q);
      sync($q, shift);
      $q.Some; send($q, x); send($q, shift);
      $q = $r;    // forward request
    case IsEmpty:
      shift = async_recv($q);
      sync($q, shift);
      send($q, false); send($q, shift);
      $q = elem(x, $r);
    case Dealloc:
      shift = async_recv($q);
      $r.Dealloc; send($r, shift);
      async_wait($r);
      sync($r, end); sync($q, shift);
      close($q); 
 } }
\end{verbatim}
\end{minipage}
&
$\quad{}$
\begin{minipage}{8cm}
\begin{verbatim}
queue $q empty () {
  switch ($q) {
    case Enq:
      int y = async_recv($q);
      queue $e = empty();
      sync($q, y);
      $q = elem(y, $e);
    case Deq:
      shift = async_recv($q);
      sync($q, shift);
      $q.None; send($q, shift);
      $q = empty();
    case IsEmpty:
      shift = async_recv($q);
      sync($q, shift);
      send($q, true); send($q, shift);
      $q = empty();
    case Dealloc:
      shift = async_recv($q);
      sync($q, shift);
      close($q);
  }
}
\end{verbatim}
\end{minipage}
\end{tabular}
\caption{Implementation of a queue with constant span enqueue and dequeue operations from client's perspective, in CC0's intermediate language using non-blocking input.}
\label{program:targetlang}
\end{program}

\subsection{Cost Semantics}

We define a receive request to increase both span and work by one unit. Upon synchronising any request, the span must also be synchronized with the value carried by the message. As before, during the synchronization, any $\m{end}$ or $\m{fwd}$ message will require the addition of the work contained in the message to the work of the receiving process.

There is also the need to change the definition of Configuration, modifying the $\m{queue}$ predicate, which will need to keep a new queue of requests, $r$, defined below.

\footnotesize

\[
\begin{array}{llcl}
\mbox{Configurations} & \Omega & ::= & \cdot \\
& & \mid & \m{queue}(\$c, q, \$d, r), \Omega \\
& & \mid & \m{proc}(\$c, P, s, w), \Omega \\
& & \mid & \m{cell}(\$c, x, v), \Omega \\
\end{array}
\]

\[
\begin{array}{l@{\hspace{1em}}lcll}
\mbox{Request queue} & r & ::= & \cdot \mid x \cdot r \mid \m{end} \mid \m{shift}
\end{array}
\]

\normalsize

The span of a non-blocking program either decreases or is the same as the corresponding blocking one. For terminating programs, work remains the same between the two models. We have a proof sketch for these two intuitive properties, but a more rigorous proof is still in development.  Here, we only show three representative rules.

\footnotesize

\[
\begin{array}{lcl}
\m{data}\_\m{async\_r} & : & \m{proc}(\$e, x = \m{async\_recv}(\$c) \semi Q, s, w) \tensor
\m{queue}(\$c, \lft{q}, \$d, r) \\
& & \quad \lolli \{\exists x.\, \m{proc}(\$e, Q, s + 1, w+1) \tensor \m{queue}(\$c, \lft{q}, \$d, r \cdot x) \} \\[1ex]
\m{sync}\_\m{wait \; 1} & : &\m{proc}(\$e, \m{sync}(\$c, \; \m{end}) \semi Q, s, w) \tensor \m{queue}(\$c, \lft{(v,s_1,w_1) \cdot q}, \$d, y \cdot r) \\
& & \quad \lolli \{ \m{proc}(\$e, \m{sync}(\$c, \; \m{end}) \semi Q, max(s,s_1), w) \tensor \m{queue}(\$c, \lft{q}, \$d, r) \tensor \m{cell}(\$e,y,v)\} \\[1ex]
\m{sync}\_\m{wait \; 2} & : &\m{proc}(\$e, \m{sync}(\$c, \; \m{end}) \semi Q, s, w) \tensor \m{queue}(\$c, \lft{(\m{end},s_1,w_1)}, \$d, \m{end}) \\
& & \quad \lolli \{ \m{proc}(\$e, Q, max(s,s_1), w + w_1)\}
\end{array}
\]

\normalsize

\section{Translation}

We represent the translation from a blocking program to a non-blocking one by $\denote{\;}$. It requires an auxiliary table $\sigma$ of requests, which is used to determine where to include the synchronization functions. This table is represented by a set of pairs, $(\$c,x)$, where $\$c$ is a channel and $x$ is either a variable, an $\m{end}$ or a $\m{shift}$. Each process has its own local table. The $\denote{\;}$ function takes a pair $(\m{Instruction},\sigma)$ and returns another pair carrying the translation of the instruction and a new table. Table \ref{table:translation} presents some examples of the translation.

\begin{table}
\footnotesize
\begin{tabular}{lcl}
$\cdot$ \denote{$(x = \m{recv}(\$d), \; \sigma)$}  & = & $(x = \m{async\_recv}(\$d), \; \sigma \cup \{(x,\$d)\})$ \\		
\end{tabular}
$\newline$
$\newline$
\begin{tabular}{lcl} 
$\cdot$ \denote{$(\m{close}(\$d), \; \sigma)$}  & = & $(sync\_instructions; \; \m{close}(\$d), \; \{\})$ \\
							       		& & \begin{tabular}{ll} where & $l = \m{sync\_all}$ $\sigma$ \\
							       		 					    & $(sync\_instructions,\sigma') = \m{generate\_sync}$ $l$ $\sigma$ \\
														    \end{tabular}						       				
\end{tabular}
$\newline$
$\newline$
\begin{tabular}{lcl} 
$\cdot$ \denote{$(\m{send}(\$d, e), \; \sigma)$}  & = & $(sync\_instructions; \; \m{send}(\$d, e), \; \sigma')$ \\
							       		& & \begin{tabular}{ll} where & $ l_1 = \m{check\_shift}$ $\sigma$ $\$d$ \\
														    & $ l_2 = \m{check\_exp}$ $\sigma$ $e$ \\
							       		 					    & $(sync\_instructions,\sigma') = \m{generate\_sync}$ $(l_1 \cup \l_2)$ $\sigma$ \\
														    \end{tabular}						       				
\end{tabular}
\caption{Translation scheme of $\m{recv}$-value, close and $\m{send}$-expression operations.}
\label{table:translation}
\end{table}

To translate a receive, there is no synchronization needed, which is a consequence of defining the rule to bind a new variable. The function $\m{async\_recv}$ produces a request so we have to include a pair in the table of requests, indicating both the variable and the channel tied to the request.

The translations for $\m{close}$ and send an expression are similar. To close a channel we need to synchronize it and all its clients, so $\m{sync\_all}$ generates a list with a number of pairs equal to the number of requests in $\sigma$. The auxiliary function $\m{generate\_sync}$ takes this list and returns the minimum sequence of instructions needed to synchronize the channel. If there is no need to perform any synchronization, $\m{generate\_sync}$ produces a no operation instruction. This behaviour is captured by the variable $sync\_instructions$: which contains a (possibly empty) list of synchronization instructions "$\m{sync}(\$d, \m{shift}) \semi  \ldots$".

To send an expression, we need to perform two different checks: if the channel has the correct polarity and if there is any variable in the expression that needs to be synchronized. The auxiliary functions $\m{check\_shift}$ and $\m{check\_exp}$ handle these two verifications, respectively. They return a list of pairs containing the variables and channels that require synchronisation, lists that are passed on to $\m{generate\_sync}$, which produces the synchronisation instructions, as mentioned previously. This function also removes all the synchronized pairs from the table of requests.

\section{Experimental Evaluation}

All benchmarks were run on a 2015 Macbook Pro, with a 2.7 GHz Intel Core I5 (2 cores) processor and 8 GB RAM. The benchmarking suite\footnote{available at \url{http://www.cs.cmu.edu/~fp/misc/cc0-bench.tgz}} can be found in table 2, adapted from \cite{willseydesign}.

\begin{table}[H]
\setlength{\tabcolsep}{8pt}
\begin{center}
\scriptsize\begin{tabular}{l p{4cm} |l p{5cm}}
bitstring1 & bitstrings with external choice & parfib & parallel naive Fibonacci, simulating
fork/join \\
bitstring3 &  bitstrings with internal choice & primes &  prime sieve (sequential) \\
bst & binary search trees, tree sort &  queue-notail & queues without tail calls \\ 
insert-sort & insertion sort using a queue & queue & queues written naturally\\
mergesort1 & mergesort with Fibonacci trees & reduce & reduce and scan on parallel sequences\\
mergesort3 & mergesort with binary trees & seg & list segments\\
mergesort4 & mergesort with binary trees, sequential merge & sieve-eager & eager prime sieve
merge\\
odd-even-sort1 & odd/even sort, v1 & sieve-lazy & lazy prime sieve\\
odd-even-sort4 & odd/even sort, v4 & stack & a simple stack\\
odd-even-sort6 & odd/even sort, v6 & & \\
\end{tabular}
\caption{CC0 benchmarking suite}
\end{center}
\label{table:bsuite}
\end{table}

\begin{figure}[H]
\includegraphics[width=\textwidth, height=5cm]{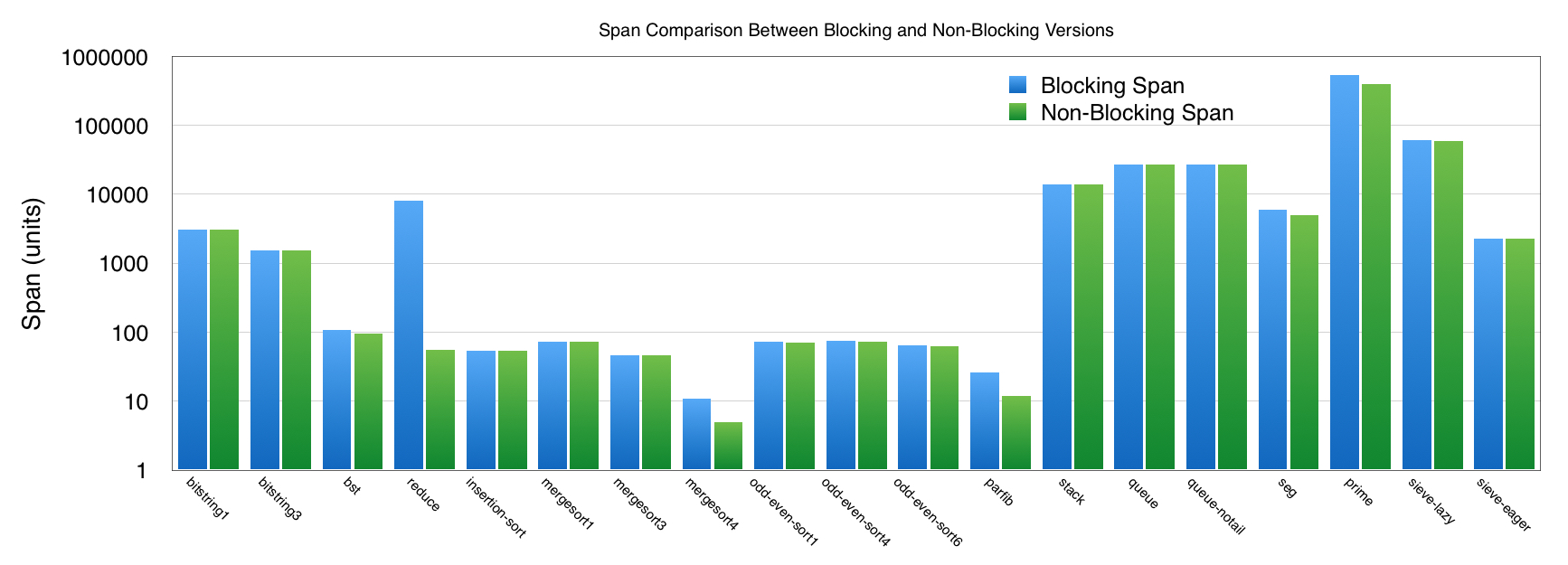}
\caption{Blocking and Non-Blocking span benchmarks on a log scale.}
\label{figure:spans}
\end{figure}

\begin{figure}[H]
\includegraphics[width=\textwidth, height=5cm]{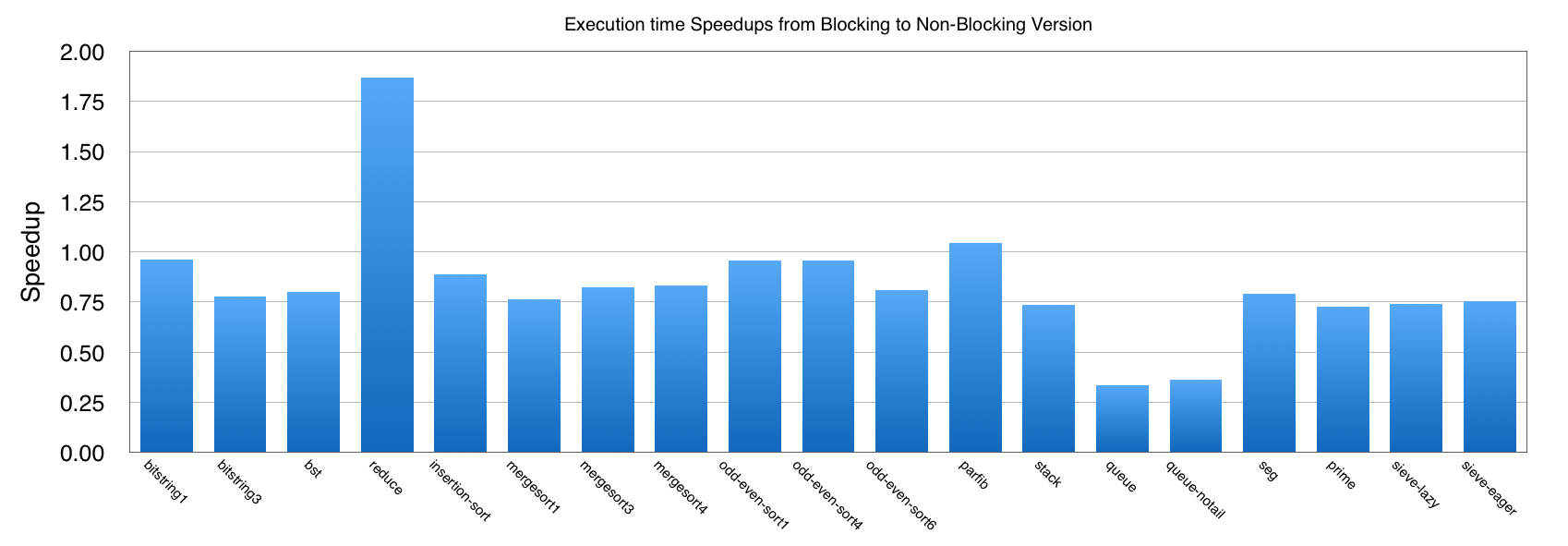}
\caption{Execution time speedup from Blocking to Non-Blocking Version.}
\label{figure:speedups}
\end{figure}

Note that the span is often the same across our benchmark suite, where $\m{reduce}$, $\m{mergesort4}$, and $\m{parfib}$ see noticeable improvements.  In the case of $\m{reduce}$, some of the difference could be recovered by fine-tuning the CC0 source program; it is interesting that our generic technique can make up for a performance bug (when considered under the blocking semantics) introduced by the programmer. This shows that, at the very least, our implementation can help identify some performance issues in the given code.  As figure~\ref{figure:speedups} shows, the overhead of maintaining the request queue is considerable in some examples.  Only in $\m{reduce}$ and $\m{parfib}$ do we realize an actual performance improvement.

\section{Future Work}

Since recipient behavior on input is opaque to the sender, we may be able to craft an optimisation which avoids the slowdown in the common case where no improvement in the span is available. For this purpose we would likely combine our fully dynamic technique with static dependency analysis to use non-blocking input only where promising. Our cost semantics could also become a tool in performance debugging which may be helpful in particular to novice programmers with little experience in concurrency.

\textbf{Acknowledgments:} This work is partially funded by the FCT (Portuguese Foundation for Science and Technology) through the Carnegie Mellon Portugal Program.

\bibliographystyle{eptcs}
\bibliography{refs}

\begin{thebibliography}{10}
\providecommand{\bibitemdeclare}[2]{}
\providecommand{\surnamestart}{}
\providecommand{\surnameend}{}
\providecommand{\urlprefix}{Available at }
\providecommand{\url}[1]{\texttt{#1}}
\providecommand{\href}[2]{\texttt{#2}}
\providecommand{\urlalt}[2]{\href{#1}{#2}}
\providecommand{\doi}[1]{doi:\urlalt{http://dx.doi.org/#1}{#1}}
\providecommand{\bibinfo}[2]{#2}

\bibitemdeclare{inproceedings}{baker1977futures}
\bibitem{baker1977futures}
\bibinfo{author}{Henry~C. \surnamestart Baker\surnameend, Jr.} \&
  \bibinfo{author}{Carl \surnamestart Hewitt\surnameend}
  (\bibinfo{year}{1977}): \emph{\bibinfo{title}{The Incremental Garbage
  Collection of Processes}}.
\newblock In: {\sl \bibinfo{booktitle}{Proceedings of the 1977 Symposium on
  Artificial Intelligence and Programming Languages}},
  \bibinfo{publisher}{ACM}, \bibinfo{address}{New York, NY, USA}, pp.
  \bibinfo{pages}{55--59}, \doi{10.1145/800228.806932}.

\bibitemdeclare{inproceedings}{balzer2015objects}
\bibitem{balzer2015objects}
\bibinfo{author}{Stephanie \surnamestart Balzer\surnameend} \&
  \bibinfo{author}{Frank \surnamestart Pfenning\surnameend}
  (\bibinfo{year}{2015}): \emph{\bibinfo{title}{Objects As Session-typed
  Processes}}.
\newblock In: {\sl \bibinfo{booktitle}{Proceedings of the 5th International
  Workshop on Programming Based on Actors, Agents, and Decentralized Control}},
  \bibinfo{series}{AGERE! 2015}, \bibinfo{publisher}{ACM},
  \bibinfo{address}{New York, NY, USA}, pp. \bibinfo{pages}{13--24},
  \doi{10.1145/2824815.2824817}.

\bibitemdeclare{inproceedings}{caires2010session}
\bibitem{caires2010session}
\bibinfo{author}{Lu{\'\i}s \surnamestart Caires\surnameend} \&
  \bibinfo{author}{Frank \surnamestart Pfenning\surnameend}
  (\bibinfo{year}{2010}): \emph{\bibinfo{title}{Session Types as Intuitionistic
  Linear Propositions}}.
\newblock In: {\sl \bibinfo{booktitle}{Proceedings of the 21st International
  Conference on Concurrency Theory (CONCUR 2010)}},
  \bibinfo{publisher}{Springer LNCS 6269}, \bibinfo{address}{Paris, France},
  pp. \bibinfo{pages}{222--236}, \doi{10.1007/978-3-642-15375-4\_16}.

\bibitemdeclare{article}{caires2013linear}
\bibitem{caires2013linear}
\bibinfo{author}{Lu{\'i}s \surnamestart Caires\surnameend},
  \bibinfo{author}{Frank \surnamestart Pfenning\surnameend} \&
  \bibinfo{author}{Bernardo \surnamestart Toninho\surnameend}
  (\bibinfo{year}{2016}): \emph{\bibinfo{title}{Linear logic propositions as
  session types}}.
\newblock {\sl \bibinfo{journal}{Mathematical Structures in Computer Science}}
  \bibinfo{volume}{26}, pp. \bibinfo{pages}{367--423},
  \doi{10.1017/S0960129514000218}.

\bibitemdeclare{article}{Cervesato09ic}
\bibitem{Cervesato09ic}
\bibinfo{author}{Iliano \surnamestart Cervesato\surnameend} \&
  \bibinfo{author}{Andre \surnamestart Scedrov\surnameend}
  (\bibinfo{year}{2009}): \emph{\bibinfo{title}{Relating state-based and
  process-based concurrency through linear logic (full-version)}}.
\newblock {\sl \bibinfo{journal}{Information and Computation}}
  \bibinfo{volume}{207}(\bibinfo{number}{10}), pp. \bibinfo{pages}{1044 --
  1077}, \doi{10.1016/j.ic.2008.11.006}.
\newblock \bibinfo{note}{Special issue: 13th Workshop on Logic, Language,
  Information and Computation (WoLLIC 2006)}.

\bibitemdeclare{inbook}{Cormen:2001:IA:580470}
\bibitem{Cormen:2001:IA:580470}
\bibinfo{author}{Thomas~H. \surnamestart Cormen\surnameend},
  \bibinfo{author}{Clifford \surnamestart Stein\surnameend},
  \bibinfo{author}{Ronald~L. \surnamestart Rivest\surnameend} \&
  \bibinfo{author}{Charles~E. \surnamestart Leiserson\surnameend}
  (\bibinfo{year}{2001}): \emph{\bibinfo{title}{Introduction to Algorithms}},
  \bibinfo{edition}{2nd} edition, chapter~\bibinfo{chapter}{27}.
\newblock \bibinfo{publisher}{McGraw-Hill Higher Education}.

\bibitemdeclare{article}{Friedman78aspectsof}
\bibitem{Friedman78aspectsof}
\bibinfo{author}{Daniel~P. \surnamestart Friedman\surnameend},
  \bibinfo{author}{\surnamestart David\surnameend} \&
  \bibinfo{author}{S.~\surnamestart Wise\surnameend} (\bibinfo{year}{1978}):
  \emph{\bibinfo{title}{Aspects of applicative programming for parallel
  processing}}.
\newblock {\sl \bibinfo{journal}{IEEE Transactions on Computers}}, pp.
  \bibinfo{pages}{289--296}, \doi{10.1109/TC.1978.1675100}.

\bibitemdeclare{unpublished}{Guenot14un}
\bibitem{Guenot14un}
\bibinfo{author}{Nicolas \surnamestart Guenot\surnameend}
  (\bibinfo{year}{2014}): \emph{\bibinfo{title}{Session Types, Solos, and the
  Computational Contents of the Sequent Calculus}}.
\newblock \bibinfo{note}{Talk at the Types Meeting}.

\bibitemdeclare{inproceedings}{honda1993types}
\bibitem{honda1993types}
\bibinfo{author}{Kohei \surnamestart Honda\surnameend} (\bibinfo{year}{1993}):
  \emph{\bibinfo{title}{Types for Dyadic Interaction}}.
\newblock In: {\sl \bibinfo{booktitle}{4th International Conference on
  Concurrency Theory}}, \bibinfo{series}{CONCUR'93},
  \bibinfo{publisher}{Springer LNCS 715}, pp. \bibinfo{pages}{509--523},
  \doi{10.1007/3-540-57208-2\_35}.

\bibitemdeclare{inproceedings}{honda1998language}
\bibitem{honda1998language}
\bibinfo{author}{Kohei. \surnamestart Honda\surnameend},
  \bibinfo{author}{Vasco~T. \surnamestart Vasconcelos\surnameend} \&
  \bibinfo{author}{Makoto \surnamestart Kubo\surnameend}
  (\bibinfo{year}{1998}): \emph{\bibinfo{title}{Language Primitives and Type
  Discipline for Structured Communication-Based Programming}}.
\newblock In: {\sl \bibinfo{booktitle}{7th European Symposium on Programming
  Languages and Systems}}, \bibinfo{series}{ESOP'98},
  \bibinfo{publisher}{Springer LNCS 1381}, pp. \bibinfo{pages}{122--138},
  \doi{10.1007/BFb0053567}.

\bibitemdeclare{article}{Laneve03mscs}
\bibitem{Laneve03mscs}
\bibinfo{author}{Cosimo \surnamestart Laneve\surnameend} \&
  \bibinfo{author}{Bj{\"o}rn \surnamestart Victor\surnameend}
  (\bibinfo{year}{2003}): \emph{\bibinfo{title}{Solos in Concert}}.
\newblock {\sl \bibinfo{journal}{Mathematical Structures in Computer Science}}
  \bibinfo{volume}{13}(\bibinfo{number}{5}), pp. \bibinfo{pages}{657--683},
  \doi{10.1017/S0960129503004055}.

\bibitemdeclare{inproceedings}{Pfenning09lics}
\bibitem{Pfenning09lics}
\bibinfo{author}{F.~\surnamestart Pfenning\surnameend} \&
  \bibinfo{author}{R.~J. \surnamestart Simmons\surnameend}
  (\bibinfo{year}{2009}): \emph{\bibinfo{title}{Substructural Operational
  Semantics as Ordered Logic Programming}}.
\newblock In: {\sl \bibinfo{booktitle}{Logic In Computer Science, 2009. LICS
  '09. 24th Annual IEEE Symposium on}}, pp. \bibinfo{pages}{101--110},
  \doi{10.1109/LICS.2009.8}.

\bibitemdeclare{misc}{c0}
\bibitem{c0}
\bibinfo{author}{Frank \surnamestart Pfenning\surnameend} \&
  \bibinfo{author}{Rob \surnamestart Arnold\surnameend} (\bibinfo{year}{2010}):
  \emph{\bibinfo{title}{C0 Language}}.
\newblock \urlprefix\url{http://c0.typesafety.net}.

\bibitemdeclare{inproceedings}{pfenning2015polarized}
\bibitem{pfenning2015polarized}
\bibinfo{author}{Frank \surnamestart Pfenning\surnameend} \&
  \bibinfo{author}{Dennis \surnamestart Griffith\surnameend}
  (\bibinfo{year}{2015}): \emph{\bibinfo{title}{Polarized Substructural Session
  Types}}.
\newblock In \bibinfo{editor}{A.~\surnamestart Pitts\surnameend}, editor: {\sl
  \bibinfo{booktitle}{Proceedings of the 18th International Conference on
  Foundations of Software Science and Computation Structures (FoSSaCS 2015)}},
  \bibinfo{publisher}{Springer LNCS 9034}, \bibinfo{address}{London, England},
  pp. \bibinfo{pages}{3--22}, \doi{10.1007/978-3-662-46678-0\_1}.

\bibitemdeclare{inproceedings}{Toninho12fossacs}
\bibitem{Toninho12fossacs}
\bibinfo{author}{Bernardo \surnamestart Toninho\surnameend},
  \bibinfo{author}{Lu{\'\i}s \surnamestart Caires\surnameend} \&
  \bibinfo{author}{Frank \surnamestart Pfenning\surnameend}
  (\bibinfo{year}{2012}): \emph{\bibinfo{title}{Functions as Session-Typed
  Processes}}.
\newblock In \bibinfo{editor}{L.~\surnamestart Birkedal\surnameend}, editor:
  {\sl \bibinfo{booktitle}{15th International Conference on Foundations of
  Software Science and Computation Structures}}, \bibinfo{series}{FoSSaCS'12},
  \bibinfo{publisher}{Springer LNCS}, \bibinfo{address}{Tallinn, Estonia}, pp.
  \bibinfo{pages}{346--360}, \doi{10.1007/978-3-642-28729-9\_23}.

\bibitemdeclare{inproceedings}{Toninho2013}
\bibitem{Toninho2013}
\bibinfo{author}{Bernardo \surnamestart Toninho\surnameend},
  \bibinfo{author}{Luis \surnamestart Caires\surnameend} \&
  \bibinfo{author}{Frank \surnamestart Pfenning\surnameend}
  (\bibinfo{year}{2013}): \emph{\bibinfo{title}{Higher-Order Processes,
  Functions, and Sessions: A Monadic Integration}}.
\newblock In: {\sl \bibinfo{booktitle}{Proceedings of the 22Nd European
  Conference on Programming Languages and Systems}}, \bibinfo{series}{ESOP'13},
  \bibinfo{publisher}{Springer-Verlag}, \bibinfo{address}{Berlin, Heidelberg},
  pp. \bibinfo{pages}{350--369}, \doi{10.1007/978-3-642-37036-6\_20}.

\bibitemdeclare{article}{wadler2012propositions}
\bibitem{wadler2012propositions}
\bibinfo{author}{Philip \surnamestart Wadler\surnameend}
  (\bibinfo{year}{2015}): \emph{\bibinfo{title}{Propositions As Types}}.
\newblock {\sl \bibinfo{journal}{Commun. ACM}}
  \bibinfo{volume}{58}(\bibinfo{number}{12}), pp. \bibinfo{pages}{75--84},
  \doi{10.1145/2699407}.

\bibitemdeclare{misc}{willseydesign}
\bibitem{willseydesign}
\bibinfo{author}{Max \surnamestart Willsey\surnameend},
  \bibinfo{author}{Rokhini \surnamestart Prabhu\surnameend} \&
  \bibinfo{author}{Frank \surnamestart Pfenning\surnameend}
  (\bibinfo{year}{2016}): \emph{\bibinfo{title}{Design and Implementation of
  Concurrent C0}}.
\newblock \urlprefix\url{https://www.cs.cmu.edu/~fp/papers/cc016.pdf}.

\end{thebibliography}

\end{document}